\documentclass[conference]{IEEEtran}
\IEEEoverridecommandlockouts
\usepackage{cite}
\usepackage{amsmath,amssymb,amsfonts}
\usepackage{algorithmic}
\usepackage{graphicx}
\usepackage{textcomp}
\usepackage{algorithm}
\usepackage{algorithmic}
\usepackage{xcolor}

\usepackage[left=1.62cm,right=1.62cm,top=0.7in]{geometry}

\newlength\myindent
\ifCLASSOPTIONcompsoc\usepackage[caption=false, font=normalsize, labelfont=sf, textfont=sf]{subfig}
\else
\usepackage[caption=false, font=footnotesize]{subfig}

\def\BibTeX{{\rm B\kern-.05em{\sc i\kern-.025em b}\kern-.08em
    T\kern-.1667em\lower.7ex\hbox{E}\kern-.125emX}}
\begin{document}
\title{On the BER vs. Bandwidth-Efficiency Trade-offs in Windowed OTSM Dispensing with Zero-Padding\\
}

\author{\IEEEauthorblockN{Zeping Sui\IEEEauthorrefmark{1}, Hongming Zhang\IEEEauthorrefmark{2}, Hien Quoc Ngo\IEEEauthorrefmark{1}, Michail Matthaiou\IEEEauthorrefmark{1} and Lajos~Hanzo\IEEEauthorrefmark{3}}
\IEEEauthorblockA{\IEEEauthorrefmark{1}Centre for Wireless Innovation (CWI), Queen's University Belfast, U.K. \\\IEEEauthorrefmark{2}School of Information and Communication Engineering,\\ Beijing University of Posts and Telecommunications, Beijing, China\\\IEEEauthorrefmark{3}School of Electronics and Computer Science, University of Southampton, Southampton, U.K.\\E-mail:\{z.sui, hien.ngo, m.matthaiou\}@qub.ac.uk, zhanghm@bupt.edu.cn, lh@ecs.soton.ac.uk}
}
\maketitle
\begin{abstract}
An orthogonal time sequency multiplexing (OTSM) scheme using practical signaling functions is proposed under strong phase noise (PHN) scenarios. By utilizing the transform relationships between the delay-sequency (DS), time-frequency (TF) and time-domains, we first conceive the DS-domain input-output relationship of our OTSM system, where the conventional zero-padding is discarded to increase the spectral efficiency. Then, the unconditional pairwise error probability is derived, followed by deriving the bit error ratio (BER) upper bound in closed-form. Moreover, we compare the BER performance of our OTSM system based on several practical signaling functions. Our simulation results demonstrate that the upper bound derived accurately predicts the BER performance in the case of moderate to high signal-to-noise ratios (SNRs), while harnessing practical window functions is capable of attaining an attractive out-of-band emission (OOBE) \emph{vs.} BER trade-off.
\end{abstract}
~
\IEEEpeerreviewmaketitle

\section{Introduction}\label{Section1}
Recently, the delay-Doppler (DD)-domain orthogonal time frequency space (OTFS) modulation scheme has emerged as an attractive waveform candidate for the 6G wireless networks since it is capable of handling high mobility scenarios \cite{7925924,9508932,8424569}. Specifically, by invoking the inverse symplectic finite Fourier transform (ISFFT)/SFFT, each DD-domain symbol is spread across the TF-domain plane, yielding the maximum achievable diversity order \cite{9404861,8686339,10217007}. The approximate orthogonalities in  delay and Doppler domains can be achieved by invoking time and frequency periodic basis functions \cite{li2023pulse}. More importantly, since the sparse DD-domain representation of doubly-selective channels can be viewed as being quasi-static \cite{steele1999mobile,9508141,raviteja2018practical}, OTFS can attain a better BER performance than conventional orthogonal frequency division multiplexing (OFDM) \cite{10129061,10250854,9891774,9439819}. 

However, the complexity of the ISFFT/SFFT becomes excessive, particularly when the number of subcarriers and time-slots is high \cite{10250854,yuan2020simple,thaj2020low}. To this end, by mapping the symbols onto the DS-domain and harnessing the inverse Walsh-Hadamard transform (IWHT) to obtain the delay-time (DT) domain signal frame, OTSM has been proposed in \cite{thaj2021orthogonal1}, where sequency denotes the number of zero-crossings per unit time. It was observed in \cite{thaj2021orthogonal} that OTSM is capable of attaining a similar BER performance to that of OTFS at a lower system complexity, since the IWHT reduces the number of multiplication operations \cite{thaj2021orthogonal,10183832}. Thaj \emph{et al.} conceived a time-domain Gauss-Seidel detector for zero-padding-aided OTSM (ZP-OTSM) systems. Later in \cite{10183832}, the input-output relationship of rectangular pulse-based non-ZP OTSM systems was derived, and a low-complexity vector approximate message passing detector was proposed. More recently, index modulation was combined with OTSM to obtain error performance gains \cite{10195188}. Nonetheless, the above-mentioned OTSM schemes all utilize rectangular transmit windows, hence their sinc-shaped spectrum results in high OOBE \cite{7120032}. Moreover, the system spectral efficiency suffers a lot due to the adoption of ZP. Furthermore, in realistic electronic circuit based scenarios, the jitter of the oscillator varies rapidly with the voltage \cite{8993708,6868950,4277101}, resulting in severe PHN, which was ignored in the above mentioned OTSM related works.
\begin{figure*}[t]
\centering
\includegraphics[width=0.8\linewidth]{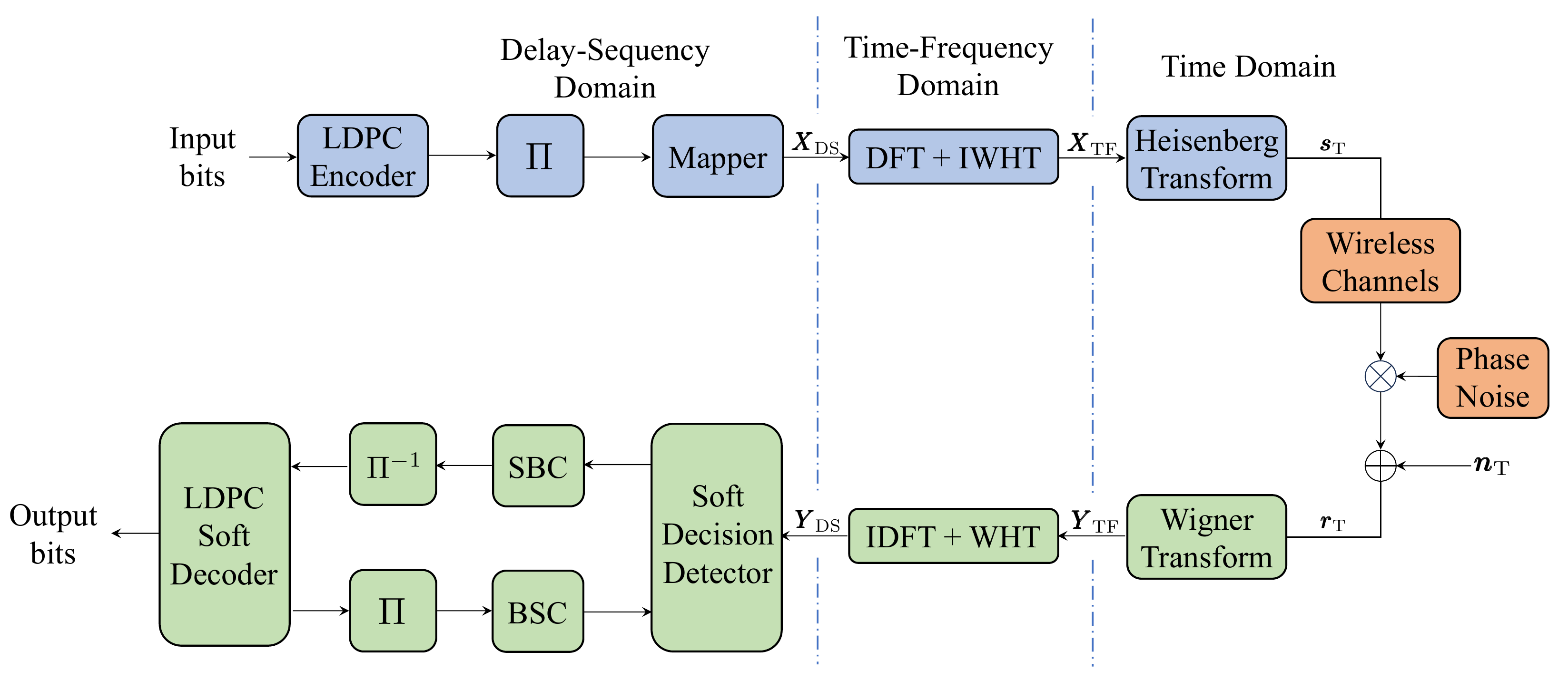}
\caption{Illustration of the LDPC-coded OTSM system with practical windows and PHN, where $\Pi$ and $\Pi^{-1}$ respresent the interleaver and deinterleaver, respectively.}
\label{Figure1}
\end{figure*}
To address the above issues, by dispensing with ZP, we conceive a practical smooth window-based band-limited OTSM system operating in the face of realistic PHN. The contributions of our paper are detailed as follows:
\begin{itemize}
	\item We first derive the DS-domain input-output relationship of the OTSM system based on the transform relationships of different domains, where both arbitrary window waveforms and the PHN effects are considered.
	\item Then, based on the input-output relationship derived, the conditional pairwise error probability (CPEP) and the unconditional PEP (UPEP) of our practical OTSM systems are derived in the presence of PHN by exploiting the pairwise error events. Furthermore, we derive a closed-form BER upper bound of our OTSM systems, yielding an accurate prediction of the BER performance at high SNRs.
\item By exploiting several practical windows functions, we compare both the OOBE and the BER performance of low-density parity-check (LDPC)-coded OTSM systems by simulations. It is demonstrated that there is a trade-off between the BER and the OOBE.
\end{itemize}

\section{System Model}\label{Section2}
\subsection{Transmitter Description of Non-ZP OTSM Relying on Smooth Windowing}
Let us consider an OTSM system having a bandwidth of $B=M\Delta f$ and frame duration of $T_f=NT$, where $M$ and $N$ represent the number of bins along the delay- and sequency-domains, while $\Delta f$ and $T=1/\Delta f$ denote the frequency-domain (FD) sampling interval and the symbol duration, respectively. As illustrated in Fig. \ref{Figure1}, the LDPC-coded information bits are mapped onto $MN$ symbols based on a $Q$-ary normalized constellation $\mathcal{A}=\{a_1,\ldots,a_Q\}$, yielding a DS-domain OTSM frame $\pmb{X}_{\text{DS}}\in\mathbb{C}^{M\times N}$. Therefore, the overall number of transmitted bits is $L=MN\log_2Q$. Consequently, upon exploiting the discrete Fourier transform (DFT) and IWHT along the delay- and sequency-domains, the elements of the TF-domain transmit frame matrix can be obtained as
\begin{align}\label{Eq1}
	X_{\text{TF}}(m,n)=\sum_{l=0}^{M-1}\sum_{k=0}^{N-1}\frac{X_{\text{DS}}(l,k)}{\sqrt{MN}}\mathcal{W}_N(n,k)e^{-j2\pi\frac{ml}{M}},
\end{align}
for $n=0,\ldots,N-1$ and $m=0,\ldots,M-1$. More specifically, we have $\mathcal{W}_N(n,k)=W(n,m/N+1/2N)/\sqrt{N}$, where $W(n,\xi)$ is the Walsh function associated with $0\leq\xi\leq1$. Therefore, \eqref{Eq1} can be rewritten in the vectorial form of $\pmb{X}_{\text{TF}}=\pmb{\mathcal{F}}_M\pmb{X}_{\text{DS}}\pmb{\mathcal{W}}_N$, where $\pmb{\mathcal{F}}_M$ and $\pmb{\mathcal{W}}_N$ denote the normalized $M$-point DFT and $N$-point WHT matrices, respectively. Then, by employing the Heisenberg transform, the transmitted time-domain (TD) signal can be expressed as \cite{10129061,8424569,9508141}
\begin{align}\label{Eq2}
	s(t)=\sum_{n=0}^{N-1}\sum_{m=0}^{M-1}X_{\text{TF}}(m,n)g_{\text{tx}}(t-nT)e^{j2\pi m\Delta f(t-nT)},	
	\end{align}
where $g_{\text{tx}}(t)$ represents the TD signaling function. Then, we sample $s(t)$ at the interval of $1/(M\Delta f)$, hence \eqref{Eq2} can be formulated in a vectorial form based on \eqref{Eq1} as
\begin{align}\label{Eq3}
	\pmb{s}=\text{vec}(\pmb{G}_\text{tx}\pmb{\mathcal{F}}_M^H\pmb{X}_{\text{TF}})=(\pmb{\mathcal{W}}_N\otimes\pmb{G}_\text{tx})\pmb{x},
\end{align}
where $\pmb{G}_\text{tx}=\text{diag}\left[g_\text{tx}(0),g_\text{tx}(T/M),\ldots,g_\text{tx}((M-1)T/M)\right]$, and $\pmb{x}=\text{vec}(\pmb{X}_\text{DS})$. Moreover, $(\cdot)^H$, $\otimes$ and $\text{vec}(\cdot)$ denote the conjugate-transpose, Kronecker product and the vectorization operators, respectively.
\subsection{PHN-contaminated Received Signals}
Next, a time-varying Rayleigh fading channel having $P$ paths is invoked, whose channel impulse response is formulated as
\begin{align}\label{Eq4}
	h(\tau,\nu)=\sum_{p=1}^{P}h_{p}\delta(\tau-\tau_p)\delta(\nu-\nu_p),
\end{align}
where $h_p$, $\tau_p$ and $\nu_p$ denote the path gain, delay- and Doppler-shifts of the $p$th channel tap, respectively, while $\delta(\cdot)$ represents the Dirac-delta function. Explicitly, upon considering a wideband system, we have $\tau_p=\frac{l_p}{M\Delta f}$ and $\nu_i=\frac{\beta_p}{NT}$ with $\beta_p=k_p+\kappa_p$. Moreover, $l_p$ and $k_p$ are the $p$th integer delay and Doppler indices, while $\kappa_p$ denote the fractional Doppler component. Note that the independent and identically distributed (i.i.d.) Gaussian variables $h_{p}$ can be formulated as $h_{p}\sim\mathcal{CN}(0,1/P)$. Let us now consider the oscillator phase noise at the receiver. Then, the received TD signal can be expressed as
\begin{align}\label{Eq5}
	r(t)=e^{j\theta(t)}\int\int h(\tau,\nu)s(t-\tau)e^{j2\pi\nu(t-\tau)}d\tau d\nu+n(t),
\end{align}
where $e^{j\theta(t)}$ and $n(t)\sim\mathcal{CN}(0,N_0)$ are the PHN at the receiver side and additive white Gaussian noise terms. By sampling \eqref{Eq5} with an interval of $1/(M\Delta f)$, the $MN\times1$-dimensional received signal vector can be formulated as
\begin{align}\label{Eq6}
	\pmb{r}=\pmb{\Theta}\pmb{H}_{\text{T}}\pmb{s}+\pmb{n}_{\text{T}},
\end{align} 
where $\pmb{\Theta}=\text{diag}[e^{j\theta(0)},e^{j\theta(1)},\ldots,e^{j\theta(MN-1)}]$ represents the PHN matrix, $\pmb{n}_\text{T}$ is the TD noise vector, and the TD channel matrix $\pmb{H}_{\text{T}}$ can be expressed as $\pmb{H}_{\text{T}}=\sum_{p=1}^Ph_p\pmb{\Pi}^{\alpha_p}\pmb{\Delta}^{\beta_p}$, where $\pmb{\Pi}$ denotes the permutation matrix, which is obtained by employing forward cyclic shift of each row of $\pmb{I}_{MN}$, while $\pmb{\Delta}=\text{diag}[z^0,\ldots,z^{MN-1}]$ with $z=e^{j\frac{2\pi}{MN}}$. Explicitly, the free-running oscillator can be modelled as a Wiener process, yielding white PHN samples as \cite{8993708,6868950}
\begin{align}\label{Eq7}
	\theta(q)=\theta(q-1)+\Delta_\text{PHN},
\end{align}
for $q=1,\ldots,MN-1$, where $\Delta_\text{PHN}$ is the difference of white PHN samples with zero-mean and a variance of $\sigma^2$. Based on the Wigner transform \cite{raviteja2018practical}, the elements of the TF-domain symbol matrix $\pmb{Y}_\text{TF}$ can be formulated as
\begin{align}\label{Eq8}
Y_\text{TF}(m,n)=\int r(t)g_{\text{rx}}^*(t-nT)e^{-j2\pi m\Delta f(t-nT)}dt,
\end{align}
where $g_{\text{rx}}(t)$ denotes the receive window function, and $(\cdot)^*$ is the conjugate operator. Therefore, \eqref{Eq8} can be rewritten in a vectorial form as $\pmb{Y}_\text{TF}=\pmb{\mathcal{F}}_M\pmb{G}_\text{rx}\pmb{R}$, where $\pmb{R}=\text{vec}^{-1}(\pmb{r})$ and $\pmb{G}_\text{rx}=\text{diag}\left[g_\text{rx}(0),g_\text{rx}(T/M),\ldots,g_\text{rx}((M-1)T/M)\right]$. By leveraging the inverse transforms of \eqref{Eq1}, the elements of the DS-domain received symbol matrix can be expressed as
\begin{align}\label{Eq9}
	Y_\text{DS}(l,k)=\sum_{m=0}^{M-1}\sum_{n=0}^{N-1}\frac{{{Y}_\text{TF}(m,n)}}{\sqrt{MN}}\mathcal{W}_N(k,n)e^{j2\pi\frac{ml}{M}},
\end{align}
for $l=0,\ldots,M-1$ and $K=0,\ldots,N-1$. It can be readily shown that $\pmb{Y}_\text{DS}=\pmb{\mathcal{F}}_M^H\pmb{Y}_{\text{TF}}\pmb{\mathcal{W}}_N=\pmb{\mathcal{F}}_M^H(\pmb{\mathcal{F}}_M\pmb{G}_\text{rx}\pmb{R})\pmb{\mathcal{W}}_N=\pmb{G}_\text{rx}\pmb{R}\pmb{\mathcal{W}}_N$, and we have 
\begin{align}\label{Eq10}	
\pmb{y}=\text{vec}(\pmb{Y}_\text{DS})=\text{vec}(\pmb{G}_\text{rx}\pmb{R}\pmb{\mathcal{W}}_N)=(\pmb{\mathcal{W}}_N\otimes\pmb{G}_\text{rx})\pmb{r}.
\end{align}
By substituting \eqref{Eq3} and \eqref{Eq6} into \eqref{Eq10}, the DS-domain input-output relationship can be formulated as
\begin{align}\label{Eq11}
	\pmb{y}&=(\pmb{\mathcal{W}}_N\otimes\pmb{G}_\text{rx})\pmb{\Theta}\pmb{H}_{\text{T}}(\pmb{\mathcal{W}}_N\otimes\pmb{G}_\text{tx})\pmb{x}+(\pmb{\mathcal{W}}_N\otimes\pmb{G}_\text{rx})\pmb{n}_{\text{T}}\nonumber\\
	&=\pmb{H}\pmb{x}+\pmb{n},
\end{align}
where $\pmb{H}=(\pmb{\mathcal{W}}_N\otimes\pmb{G}_\text{rx})\pmb{\Theta}\pmb{H}_{\text{T}}(\pmb{\mathcal{W}}_N\otimes\pmb{G}_\text{tx})$ is the DS-domain channel matrix, and we have $\pmb{n}\sim\mathcal{CN}(\pmb{0},N_0\pmb{I}_{MN})$, leading to the SNR per symbol as $\gamma=1/N_0$.
\section{Performance Analysis and Smooth Windows}\label{Section3}
\subsection{Analysis of Error Performance}\label{Section3-1}
By collecting all the channel gain coefficients $h_p$ in a vector, \eqref{Eq11} can be rewritten as
\begin{align}\label{Eq12}
	\pmb{y}=\pmb{\Sigma}(\pmb{x})\pmb{h}+\pmb{{n}},
	\end{align}
where we have $\pmb{h}=[h_1,\ldots,h_P]^T$, while the codeword matrix $\pmb{\Sigma}(\pmb{x})\in\mathbb{C}^{MN\times P}$ can be formulated as
\begin{align}\label{Eq13}
	\pmb{\Sigma}(\pmb{x})=[\underbrace{\pmb{\Upsilon}_1\pmb{x}}_{MN\times 1}\quad\pmb{\Upsilon}_2\pmb{x}\quad \ldots\quad\pmb{\Upsilon}_P \pmb{x}],
	\end{align}
where $\pmb{\Upsilon}_p=(\pmb{\mathcal{W}}_N\otimes\pmb{G}_\text{rx})(\pmb{\Theta}\pmb{\Pi}^{\alpha_p}\pmb{\Delta}^{\beta_p})(\pmb{\mathcal{W}}_N\otimes\pmb{G}_\text{tx})$, for $p=1,\ldots,P$. Here, we consider the pairwise error event $\{\pmb{x}\rightarrow\tilde{\pmb{x}}\}$, where $\pmb{x}$ and $\tilde{\pmb{x}}$ represent the transmitted and the received error symbol vectors, respectively, yielding the error symbol $\pmb{e}=\pmb{x}-\tilde{\pmb{x}}$. Therefore, the corresponding Euclidean distance is given by
\begin{align}\label{Eq14}
\eta=||\pmb{\Sigma}(\pmb{e})\pmb{h}||_2^2=\pmb{h}^H\pmb{\Xi}\pmb{h},
\end{align}
where $||\cdot||_2$ denotes the $\ell_2$-norm operator, and $\pmb{\Xi}=\pmb{\Sigma}(\pmb{e})^H\pmb{\Sigma}(\pmb{e})$ can be further expressed based on \eqref{Eq13} and \eqref{Eq14} as
\begin{align}\label{Eq15}
\pmb{\Xi}=\begin{bmatrix}
	\pmb{e}^H\pmb{\Upsilon}_1^H\pmb{\Upsilon}_1\pmb{e} & \cdots & \pmb{e}^H\pmb{\Upsilon}_1^H\pmb{\Upsilon}_P\pmb{e}\\
	\vdots & \ddots & \vdots \\
	\pmb{e}^H\pmb{\Upsilon}_P^H\pmb{\Upsilon}_1\pmb{e} & \cdots & \pmb{e}^H\pmb{\Upsilon}_P^H\pmb{\Upsilon}_P\pmb{e}	
	\end{bmatrix}.
	\end{align}
In the case of exploiting the optimal maximum likelihood detector (MLD), the CPEP can be formulated based on the Chernoff bound as \cite{zhang2016compressed,10183832}
\begin{align}\label{Eq16}
	P(\pmb{x}\rightarrow\tilde{\pmb{x}}|\pmb{h})\leq\frac{1}{2}\exp\left(-\frac{\eta}{4N_0}\right).
	\end{align}
Since $\pmb{\Xi}$ is a positive semidefinite Hermitian matrix, its rank satisfies $1\leq r\leq P$. By sorting the eigenvalues in descending order, the eigenvalues and the corresponding eigenvectors can be expressed as $\left\{\lambda_1,\ldots,\lambda_r\right\}$ and $\left\{\pmb{\varphi}_1,\ldots,\pmb{\varphi}_r\right\}$, respectively. Therefore, the CPEP upper bound can be modified as
\begin{align}\label{Eq17}
P(\pmb{x}\rightarrow\tilde{\pmb{x}}|\pmb{h})\leq\frac{1}{2}\exp\left(-\frac{\sum_{i=1}^r\lambda_i|\tilde{h}_i|^2}{4N_0}\right),
\end{align}
where $\tilde{h}_i=\left\langle\pmb{h},\pmb{\varphi}_i\right\rangle$ for $i=1,\ldots,r$, here $\left\langle\cdot,\cdot\right\rangle$ denotes the inner product between two vectors. Furthermore, it can be shown that the independent complex-valued random variables $\{\tilde{h}_1,\ldots,\tilde{h}_r\}$ have a mean of $\varsigma_i=\left\langle\mathbb{E}[\pmb{h}],\pmb{\varphi}_i\right\rangle$ and a variance of $1/(2P)$ per real dimension, when $\mathbb{E}[\cdot]$ represents the expectation operator. Therefore, based on the Ricean distribution properties, $|\tilde{h}_i|$ are Ricean distributed variables with a Ricean factor of $\xi_i=|\varsigma_i|^2$ \cite{tarokh1998space}, hence its probability density function can be written as
\begin{align}\label{Eq18}
p(|\tilde{h}_i|)=2P|\tilde{h}_i|\exp(-P|\tilde{h}_i|^2-P\xi_i)I_0(2P|\tilde{h}_i|\sqrt{\xi_i}),
\end{align}
where $I_0(\cdot)$ denotes the zero-order modified Bessel function of the first kind. Based on \eqref{Eq17} and \eqref{Eq18}, the UPEP can be formulated as \cite{tarokh1998space}
\begin{align}\label{Eq19}
P(\pmb{x}\rightarrow\tilde{\pmb{x}})\leq\frac{1}{2}\prod_{i=1}^r \frac{1}{1+\frac{\lambda_i}{4PN_0}}\exp\left(-\frac{\frac{\xi_i\lambda_i}{4PN_0}}{1+\frac{\lambda_i}{4PN_0}}\right).
\end{align}

It can be observed that $|h_i|$ follows the Rayleigh distribution in the case of $\xi_i=0$. Hence, in high SNR scenarios ($N_0\ll1$), the UPEP of \eqref{Eq19} can be rewritten as
\begin{align}\label{Eq20}
	P(\pmb{x}\rightarrow\tilde{\pmb{x}})\leq\frac{1}{2\prod_{i=1}^r\lambda_i}\left(\frac{1}{4PN_0}\right)^{-r}.
\end{align}

Finally, by invoking the union bounding technique, the BER of OTSM systems in the presence of practical windows and PHN can be approximated as
\begin{align}\label{Eq21}
P_\text{B}\approx\frac{1}{L2^{L}}\sum_{\pmb{x}}\sum_{\tilde{\pmb{x}}}P(\pmb{x}\rightarrow\tilde{\pmb{x}})d(\pmb{x},\tilde{\pmb{x}}),
\end{align}
where $d(\pmb{x},\tilde{\pmb{x}})$ is the number of different bits between $\pmb{x}$ and $\tilde{\pmb{x}}$. Consequently, the BER upper bound at high SNRs can be further reformulated based on \eqref{Eq20} and \eqref{Eq21}   as
\begin{align}\label{Eq22}
P_\text{B}\leq\frac{1}{L2^{L+1}}\sum_{\pmb{x}}\sum_{\tilde{\pmb{x}}}\frac{1}{\prod_{i=1}^r\lambda_i}\left(\frac{1}{4PN_0}\right)^{-r}d(\pmb{x},\tilde{\pmb{x}}).
\end{align}
\subsection{Smooth Windows}\label{Section3-2}
The most commonly invoked practical windows are the rectangular, Hamming, Hanning, Blackman and Bartlett-Hann waveforms \cite{5753092,1455106,21693}. Given the time-duration of the window $T_0=(M-1)T/M$ and the notation of
\begin{align}
	\Lambda(t/T)=
	\begin{cases} 
1,  & \mbox{$|t|\leq T/2$}, \\
0, & \text{otherwise,}
\end{cases}
\end{align}
the window functions can be expressed as follows \cite{1455106}:
\paragraph{Rectangular Window}
\begin{align}
	g(t)=\Lambda\left(\frac{t-T_0/2}{T_0}\right).
\end{align}
\paragraph{Hamming Window}
\begin{align}
	g(t)=\left(0.54-0.46\cos\frac{2\pi t}{T_0} \right)\Lambda\left(\frac{t-T_0/2}{T_0}\right).
\end{align}
\paragraph{Hanning Window}
\begin{align}
	g(t)=\left(0.5-0.5\cos\frac{2\pi t}{T_0} \right)\Lambda\left(\frac{t-T_0/2}{T_0}\right).
\end{align}
\paragraph{Blackman Window}
\begin{align}
	g(t)=\left(0.42-0.5\cos\frac{2\pi t}{T_0}+0.08\cos\frac{4\pi t}{T_0} \right)\Lambda\left(\frac{t-T_0/2}{T_0}\right).
\end{align}
\paragraph{Bartlett-Hann Window}
\begin{align}
	g(t)=&\left(0.62-0.48\left|t'\right|+0.38\cos 2\pi t' \right)\Lambda\left(t'\right),
\end{align}
where $t'={(t-T_0/2)}/{T_0}$.
\section{Simulation Results}\label{Section4}
The numerical results characterizing both our BER upper bound, the OOBE as well as BER performance of the OTSM systems in the presence of PHN are provided in this section.

The complexity of the MLD and calculating \eqref{Eq22} is excessive, when $MN$ is high. Therefore, we first consider a limited-dimensional OTSM system to characterize our mathematical analysis of Section \ref{Section3-1}. The maximum value of delay indices and the number of time-slots employed are $M=4$ and $N=2$. The modulation order, FD sampling interval and carrier frequency are $Q=2$, $\Delta f=15$ kHz and $f_c=40$ GHz, respectively. The maximum velocity is set to $v=500$ km/h, while the number of channel paths is $P=2$. The channel gain coefficients are given by $h_i\sim\mathcal{CN}(0,1/P)$, and the maximum normalized delay and sequency indices are $l_\text{max}=M-1$ and $k_\text{max}=N-1$, respectively \cite{8686339}. Furthermore, the delay and Doppler indices of the $i$th path obey uniform distributions, which can be formulated as $k_i\in U[-k_\text{max},k_\text{max}]$ and $l_i\in U[0,l_\text{max}]$ ($l_1=0$), respectively, while we have $l_p,k_p\in U[-1/2,1/2]$. The white PHN parameters are chosen as $\theta(0)\in U[0,360^\circ)$ and $\sigma^2=0.3^\circ$, respectively.

\begin{figure}[htbp]
\centering
\includegraphics[width=\linewidth]{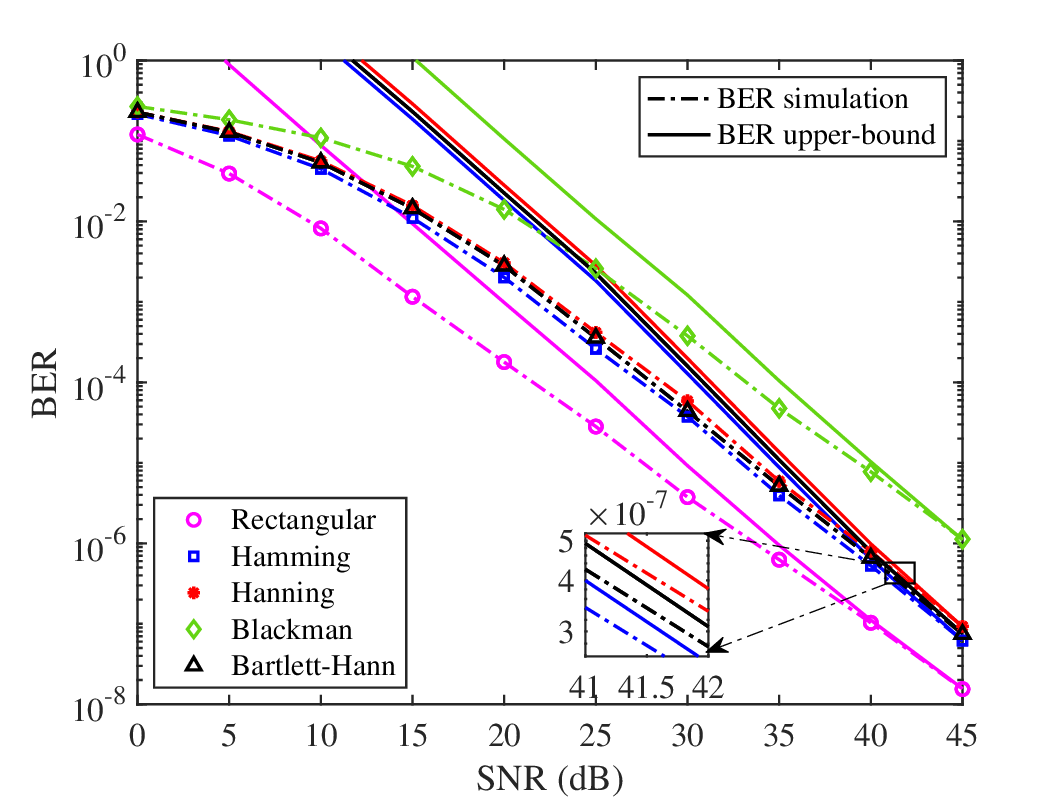}
\caption{BER performance and the upper bound of OTSM systems using different windows.}
\label{Figure2}
\end{figure}
Figure \ref{Figure2} investigates the BER performance of the MLD and compares it to the BER upper bound derived, while exploiting different practical windows, which are detailed in Section \ref{Section3-2}. It can be observed that all the theoretical upper bounds become tighter as the SNRs escalate, regardless of the choice of windows invoked. Moreover, the rectangular window-based system attains the best BER performance, followed by the Hamming, Bartlett-Hann and Hanning windows, while the BER performance of the OTSM system using Blackman window is the worst. This is because the practical windows degrade the channel gains, as seen from \eqref{Eq11}. At a BER of $10^{-6}$, the rectangular window-based OTSM system is capable of attaining about $5$ dB and $10$ dB SNR gains compared to the Hamming and Blackman scenarios, respectively. Furthermore, the BER performance and the upper bounds with the Hamming, Bartlett-Hann and Hanning windows are very close, since their mathematical expressions and shapes are similar \cite{1455106}.
\begin{figure}[htbp]
\centering
\includegraphics[width=\linewidth]{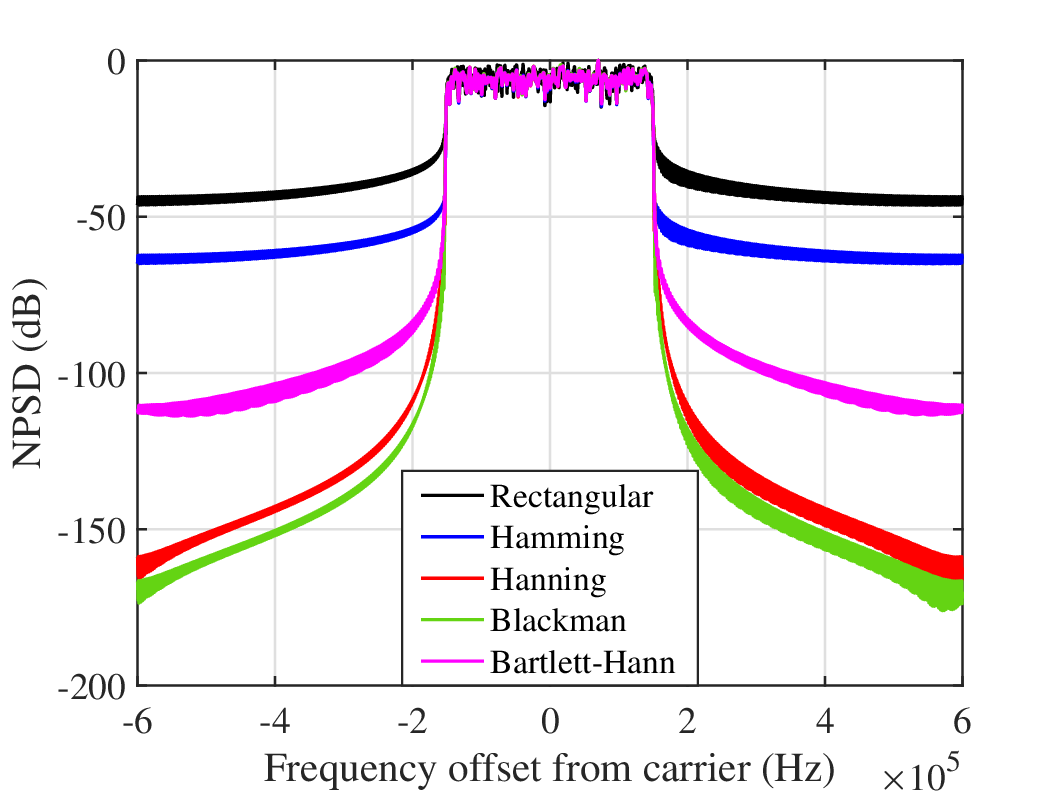}
\caption{OOBE performance of different window functions based OTSM systems.}
\label{Figure3}
\end{figure}

In Fig. \ref{Figure3}, the OOBEs of the above-mentioned windowing-based OTSM systems are plotted, where the normalized power spectral density (NPSD) is harnessed as our performance metric. Moreover, we exploit the parameter settings of ``$M=N=16, Q=4, f_c=4$ GHz, $\Delta f=18.75$ kHz'', while the maximum speed is $v_\text{max}=800$ km/h and the maximum number of sequency index are set to $l_\text{max}=6$, and the doubly-selective channel has $P=6$ paths. The remaining parameters are the same as those employed in Fig. \ref{Figure2}. It is demonstrated that some of the windows considered are capable of achieving an up to $120$ dB OOBE reduction compared to the rectangular window, at the cost of an excess bandwidth. Specifically, the Hamming and Bartlett-Hann windows achieve about $12$ dB and $60$ dB NPSD improvements compared to the rectangular windowing scenario. Moreover, it can be concluded from Fig. \ref{Figure2} and Fig. \ref{Figure3} that there is a trade-off between the BER and the OOBE of different windows.
\begin{figure}[htbp]
\centering
\includegraphics[width=\linewidth]{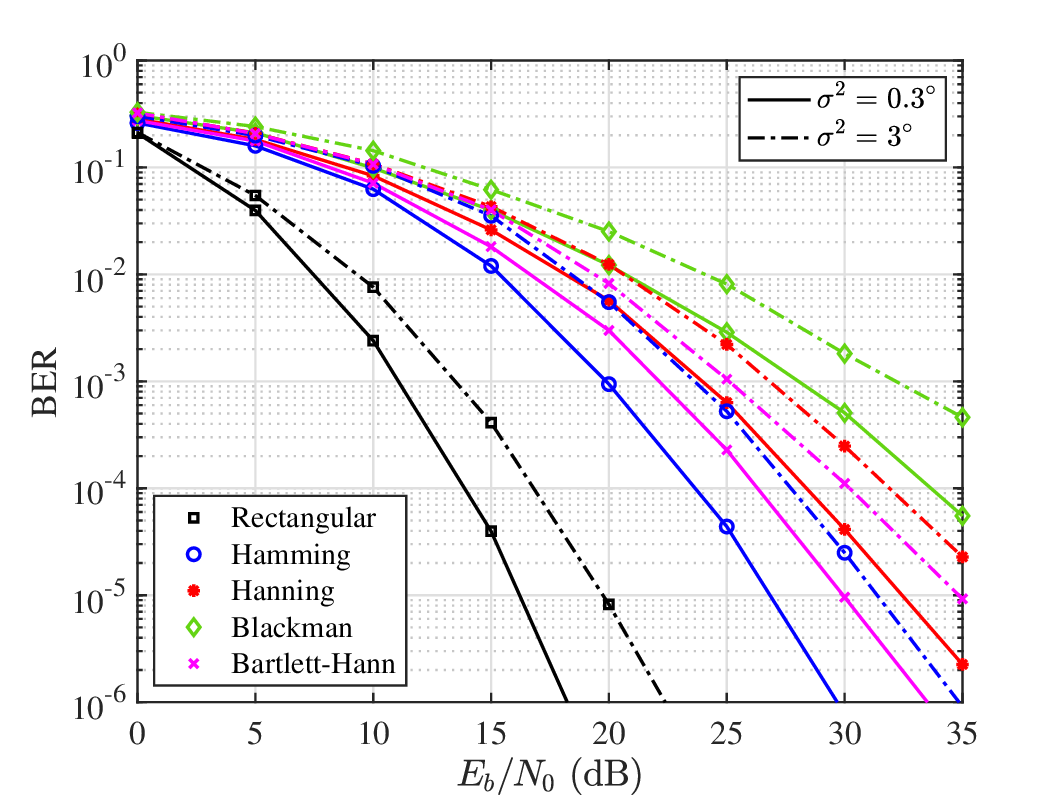}
\caption{BER performance of different windows for OTSM systems invoking an LMMSE detector in the face of white phase noise.}
\label{Figure4}
\end{figure}

To demonstrate the attractive flexibility of the OTSM system, in Fig. \ref{Figure4} we further characterize the BER performance of the corresponding $1/2$-rate LDPC-coded systems employing a soft-decision linear minimum mean square error (LMMSE) detector, symbol-to-bit converter (SBC), bit-to-symbol converter (BSC), deinterleaver, interleaver and soft LDPC decoder. Further details concerning LDPC-coded OTSM systems can be found in \cite{10183832}. Moreover, we consider two different $\sigma^2$ scenarios, while the remaining parameters are consistent with those in Fig. \ref{Figure3}. We employ $T_{\text{LDPC}}=6$ inner LDPC iterations before the soft-decoding (SD) information is passed back to the OTSM detector, this is then repeated $T_{\text{det}}=8$ times. The LDPC codeword length of $5000$, and the sum-product algorithm (SPA) are invoked \cite{910577}. Observe from Fig. \ref{Figure4} that similar to Fig. \ref{Figure2} the rectangular window-based scenario still exhibits the best BER performance at the cost of excessive OOBE. At the target BER of $10^{-6}$ associated with $\sigma^2=0.3^\circ$, the rectangular window exhibits about $12$ dB and $16$ dB SNR gains over the Hamming and Barlett-Hann windows, respectively, but the Blackman and Hanning windowing almost completely eliminates the OOBE. Moreover, it becomes explicit that the BER performance degrades substantially, when a higher value of $\sigma^2=3^\circ$ is encountered. Finally, observe from Fig. \ref{Figure2}, Fig. \ref{Figure3} and Fig. \ref{Figure4} that a flexible BER \emph{vs.} OOBE trade-off can be struck by the choice of the windowing function, since low-OOBE smooth windows can achieve lower inter-symbol interference (ISI), yielding better BER performance.
\section{Conclusions}\label{Section5}
By considering the transforms between DS- and TF-domains, we have derived the DS-domain input-output relationship of the OTSM system, which invokes arbitrary windowing functions in the presence of strong PHN. Then, we have provided the error performance analysis of our practical OTSM systems. It was demonstrated that the BER upper bound derived becomes tight at moderate to high SNRs. Furthermore, the simulation results have also illustrated that our OTSM systems relying on practical windows can strike a trade-off between the OOBE and BER. Similar to \cite{9829188}, we will exploit optimization algorithms to design optimal smooth windows to attain good BER and OOBE performance simultaneously in our future work.
\section*{Acknowledgement}
This work is a contribution by Project REASON, a UK Government funded project under the Future Open Networks Research Challenge (FONRC) sponsored by the Department of Science Innovation and Technology (DSIT). The work of M. Matthaiou was in part by the U.K. Engineering and Physical Sciences Research Council (EPSRC) under Grant EP/X04047X/1 and by the European Research Council (ERC) under the European Union's Horizon 2020 research and innovation programme (grant agreement No. 101001331). The work of H. Zhang was supported by the National Natural Science Foundation of China under Grant 62001056. L. Hanzo is the corresponding author.
\vspace{-0.2cm}
\bibliographystyle{IEEEtran}
\bibliography{IEEEabrv,OTSM_practical}

\end{document}